\documentclass[12pt]{article}
\usepackage{amsmath}
\usepackage{amssymb}
\usepackage{amsthm}
\usepackage{graphicx}
\usepackage{enumerate}
\usepackage{natbib}
\usepackage{url} 
\usepackage{multirow}
\usepackage{cellspace}
\usepackage{makecell}
\usepackage[flushleft]{threeparttable}
\usepackage{subfigure}
\usepackage{xcolor}
\newcommand{\blind}{1}

  \theoremstyle{remark}
  
\theoremstyle{plain}

 \theoremstyle{definition}
  
  \theoremstyle{plain}

\usepackage{booktabs,calc}

\providecommand{\examplename}{Example}
  \providecommand{\lemmaname}{Lemma}
  \providecommand{\remarkname}{Remark}
\providecommand{\corollaryname}{Corollary}
\providecommand{\theoremname}{Theorem}
\providecommand{\propositionname}{Proposition}

\allowdisplaybreaks[2]
\begin{document}
 \date{}

\def\spacingset#1{\renewcommand{\baselinestretch}%
{#1}\small\normalsize} \spacingset{1}


\if1\blind
{
  \title{\bf  A network-based regression approach for identifying subject-specific driver mutations}
  \author{Kin Yau Wong\hspace{.2cm}\\
    Department of Applied Mathematics, The Hong Kong Polytechnic University\\
    \\
    Donglin Zeng and D. Y. Lin \\
    Department of Biostatistics, University of North Carolina at Chapel Hill}
  \maketitle
} \fi

\spacingset{1.3}
\if0\blind
{
  \bigskip
  \bigskip
  \bigskip
  \begin{center}
    {\LARGE\bf A network-based regression approach for identifying subject-specific driver mutations}
\end{center}
  \medskip
} \fi

\bigskip
\begin{abstract}
In cancer genomics, it is of great importance to distinguish driver
mutations, which contribute to cancer progression, from causally neutral passenger
mutations.
We propose a random-effect regression approach to estimate the effects of mutations
on the expressions of genes in tumor samples, where the estimation
is assisted by a prespecified gene network. The model allows the mutation effects to vary across subjects. We develop a subject-specific mutation score to quantify the effect of a mutation
on the expressions of its downstream genes, so mutations with large scores can be prioritized as drivers. We demonstrate the usefulness of the proposed methods by simulation studies and provide an application to a breast cancer genomics study.
\end{abstract}

\noindent%
\vfill

\newpage
\spacingset{1.9} 
\section{Introduction}
\label{sec:intro}
Cancer is caused by progressive accumulation of somatic mutations. For better understanding of the disease mechanisms and
the development of effective therapeutic methods, it is of great importance to identify genetic
events that lead to cancer initiation and progression. This is a challenging task, because the genome of the tumor cells typically
harbours a large number of passenger mutations, which are causally neutral to cancer progression and need to be distinguished from
 the driver mutations that drive the progression of cancer.

Another major challenge in driver gene identification is that cancer is a highly heterogeneous disease. The same type of cancer can be driven
by different sets of mutation events. It is difficult to identity driver mutations merely based on mutation frequency, because some
less frequently mutated genes may still be drivers in tumors where they are mutated. Also, mutations in the same gene may have different
effects in different tumor samples.

One class of methods for driver gene identification incorporates the effects of mutations on other omic features to prioritize driver mutations. For example, \cite{bashashati2012drivernet} proposed DriverNet, which identifies driver mutations based on their estimated effects on mRNA expression networks.
\cite{hou2014dawnrank} proposed DawnRank, which is a computational method that assigns a score to
each gene based on the differential expression values of genes in its downstreams using a PageRank algorithm; the downstreams of a gene are given by a prespecified gene network. The method yields a subject-specific score for each gene, and mutated genes that receive high scores are regarded as drivers for a given subject. Other methods that identify driver mutations based on the impact on the expression of downstream genes include \cite{shi2016discovering}, \cite{wei2016lndriver}, and \cite{song2019random}.

The networks used in the analyses typically
encode associations discovered in a wide range of studies, and any given link in the network may not be present in the current population of interest. Also, the strength
of association among genes is not encoded in the network and thus is not taken into consideration. For example, in DawnRank, the score for a mutation is essentially a weighted sum of the differential expressions
of its downstream genes, and the weights depend only on the network structure. Therefore, the results may be sensitive to false positive links in the network and do not take into account the strength of association between the mutations and gene expressions.

We proposed a two-step approach for evaluating driver mutations that have impacts on the expressions of downstream genes. In the first step, we
fit a random effect linear model of gene expressions versus mutations, where a mutation is only allowed to have effects on the expressions of
the downstream genes specified in the network. We allow the mutation effect to vary across subjects to allow for heterogeneity. In the second step,
we evaluate the estimated subject-specific expected effect of the mutations on gene expressions given the observed data and use it as a score to
quantity how ``active'' or how important a driver each mutation is.

\section{Methods}
\label{sec:method}
Let $q$ be the number of gene expressions, $p$ be the number of mutations, $\boldsymbol{Y}\equiv(Y_1,\ldots,Y_q)^{\mathrm{T}}$ be a vector of gene expressions, and $\boldsymbol{X}\equiv(X_1,\ldots,X_p)^{\mathrm{T}}$ be a vector of mutations, with value 1 if the $j$th gene is mutated and value 0 if otherwise. In general, $\boldsymbol{X}$ and $\boldsymbol{Y}$ may correspond to the same or different sets of genes. We assume that for $k=1,\ldots,q$,
\[
Y_k=\mu_k+\sum_{j=1}^p\beta_{jk}(1+\tau b_{j})X_{j}+\epsilon_k,
\]
where $\mu_k$ is an intercept, $\beta_{1k},\ldots,\beta_{pk}$ are regression parameters, $b_1,\ldots,b_p$ are independent standard normal random variables, $\epsilon_k\sim\mathrm{N}(0,\sigma^2_k)$, $\sigma^2_k$ is a positive variance parameter, and $\tau$ is a positive parameter that characterizes the magnitude of the random effects $b_1,\ldots,b_p$. We assume that $\epsilon_1,\ldots,\epsilon_q$ are independent. In the model, the effect of the $j$th mutation on the $k$th gene expression is composed on a fixed effect $\beta_{jk}$ and a random effect $\tau\beta_{jk}b_j$, where the direction and magnitude of the random effect are characterized by the fixed effect $\beta_{jk}$. We can understand $b_j$ as a characterization of how ``active'' the $j$th mutation is, such that when $b_j$ is larger, the effect of this mutation on each expression departs from the population average value $\beta_{jk}$ by a larger extent. For a sample of size $n$, the observed data consist of $(\boldsymbol{Y}_i,\boldsymbol{X}_i)_{i=1,\ldots,n}$.

When $p$ and $q$ are large, maximum likelihood estimation is highly challenging, if not impossible. To accommodate high-dimensional data in model estimation, we propose to adopt a pseudo-likelihood approach and estimate $\beta_{jk}$'s based on the likelihood without random effects, and impose sparsity on $\beta_{jk}$'s. We define the (unpenalized) pseudo-likelihood to be
\[
\widetilde{L}(\boldsymbol{\xi})=\prod_{i=1}^n\prod_{k=1}^q\frac{1}{\sigma_k}\exp\bigg\{-\frac{1}{2\sigma^2_k}\Big(Y_{ik}-\mu_k-\sum_{j=1}^pX_{ij}\beta_{jk}\Big)^2\bigg\},
\]
where $\boldsymbol{\xi}=(\boldsymbol{\mu},\boldsymbol{B},\boldsymbol{\Sigma})$, $\boldsymbol{B}^{\mathrm{T}}=(\beta_{jk})_{j=1,\ldots,p;k=1,\ldots,q}$, and $\boldsymbol{\Sigma}=\mathrm{diag}(\sigma_1^2,\ldots,\sigma^2_q)$. This is the likelihood function under $b_j=0$ for $j=1,\ldots,p$.

To construct the penalty term, we incorporate a prespecified directed gene network that informs the structure of the regression parameter matrix $\boldsymbol{B}$. We impose the following assumptions on $\boldsymbol{B}$. First, the effects of mutations should be constrained by the network structure; a mutation cannot affect the expression of a gene that is not its direct or indirect downstream. Second, the mutation effects should be sparse, such that only a few mutations have effect on the expression
of any genes. Third, the effects of mutations should be generally more sparse on genes that are further away in the network.

To achieve these three properties, we introduce the following adaptive $L_1$-penalized pseudo log-likelihood function:
\[
p\ell(\boldsymbol{\xi})=\log\widetilde{L}(\boldsymbol{\xi})-\lambda\sum_{j=1}^p\sum_{k=1}^q\theta^{-m(j,k)}w_{jk}|\beta_{jk}|,
\]
where $\lambda$ is a tuning parameter that controls the overall strength of penalty, $\theta$ is a tuning parameter with value from $(0,1)$, $m(j,k)$ is the length of the shortest path from $j$ to $k$ in the prespecified network, and $w_{jk}$ is a weighting term obtained from an initial estimation step; we set $m(j,k)=\infty$ if there is no directed path from gene $j$ to gene $k$. In this formulation, if gene $j$ is further away from gene $k$, then $m(j,k)$ is larger, and the penalty on $\beta_{jk}$ is stronger. The tuning parameter $\theta$ controls how quickly the penalty strength increases as $m(j,k)$ increases. When the dimension $p$ is small or moderate, we can set the weighting term $w_{jk}$ to be inversely proportional to the absolute value of the unpenalized maximum pseudo-likelihood estimator. Let $\widetilde{\beta}_{jk}$ be the corresponding element of the maximizer of $\widetilde{L}(\boldsymbol{\xi})$, where in the maximization $\beta_{jk}$ is fixed at 0 if $m(j,k)=\infty$. We can set $w_{jk}=|\widetilde{\beta}_{jk}|^{-1}$. When $p$ is too large, we can set $\widetilde{\beta}_{jk}$ to be the estimator of the marginal effect of $X_j$ on $Y_k$ or simply set $w_{jk}=1$. With specified values of $\lambda$, $\theta$, and $w_{jk}$'s, we maximize $p\ell(\boldsymbol{\xi})$ and obtain a generally sparse estimator for $\boldsymbol{B}$. Because the penalized estimator is generally biased, we refit the selected model by maximizing $\widetilde{L}(\boldsymbol{\xi})$ with the zero parameters from the penalized estimators fixed at zero. Let $\widehat{\boldsymbol{B}}$ denote the refitted estimator of $\boldsymbol{B}$.

With $\boldsymbol{B}$ fixed at $\widehat{\boldsymbol{B}}$, we then estimate $\tau$ and also update the estimators of $\boldsymbol{\mu}$ and $\boldsymbol{\Sigma}$. The likelihood function that incorporates the random effects is
\[
L(\tau,\boldsymbol{\mu},\boldsymbol{\Sigma};\widehat{\boldsymbol{B}})=\prod_{i=1}^n\int\prod_{k=1}^q\frac{1}{\sigma_k}\exp\bigg\{-\frac{1}{2\sigma^2_k}\Big(\widetilde{Y}_{ik}-\mu_k-\tau\sum_{j=1}^pX_{ij}\widehat{\beta}_{jk}b_{ij}\Big)^2\bigg\}e^{-\frac{1}{2}\Vert\boldsymbol{b}_i\Vert^2}\,\mathrm{d}\boldsymbol{b}_i,
\]
where $\widetilde{Y}_{ik}=Y_{ik}-\sum_{j=1}^pX_{ij}\widehat{\beta}_{jk}$, and $\boldsymbol{b}_i=(b_{i1},\ldots,b_{ip})^{\mathrm{T}}$. We compute the maximizer of $L(\tau,\boldsymbol{\mu},\boldsymbol{\Sigma};\widehat{\boldsymbol{B}})$ using the EM algorithm, with $\boldsymbol{b}_i$'s treated as missing data. The complete-data log-likelihood is
\[
\sum_{k=1}^q\bigg\{-\frac{n}{2}\log\sigma^2_k-\frac{1}{2\sigma^2_k}\sum_{i=1}^n
\Big(\widetilde{Y}_{ik}-\mu_k-\tau\sum_{j=1}^pX_{ij}\widehat{\beta}_{jk}b_{ij}\Big)^2\bigg\}.
\]
This is the log-likelihood function for a linear regression model with  heteroscedasticity. In the M-step, we first update $\boldsymbol{\mu}$ and $\tau$ using the weighted least-squares formula at the current $\boldsymbol{\Sigma}$. Let $\boldsymbol{Q}$ be a $(q+1)\times(q+1)$ symmetric matrix with upper-triangular block elements of
\[
\boldsymbol{Q}=\left(\begin{array}{cc}
n\boldsymbol{\Sigma}^{-1} & \boldsymbol{\Sigma}^{-1}\widehat{\boldsymbol{B}}\sum_{i=1}^{n}\mathrm{diag}(\boldsymbol{X}_{i})\widehat{\mathrm{E}}\boldsymbol{b}_{i}\\
 & \mathrm{tr}\Big\{\sum_{i=1}^{n}\mathrm{diag}(\boldsymbol{X}_{i})\widehat{\mathrm{E}}\boldsymbol{b}_{i}\boldsymbol{b}_{i}^{\mathrm{T}}\mathrm{diag}(\boldsymbol{X}_{i})\widehat{\boldsymbol{B}}^{\mathrm{T}}\boldsymbol{\Sigma}^{-1}\widehat{\boldsymbol{B}}\Big\}
\end{array}\right),
\]
and let
\[
\boldsymbol{u}=\Big(\begin{array}{cccc}
\sigma_{1}^{-2}\sum_{i=1}^{n}\widetilde{Y}_{i1}, & \ldots, & \sigma_{q}^{-2}\sum_{i=1}^{n}\widetilde{Y}_{iq}, & \sum_{i=1}^{n}\sum_{j=1}^{q}\sigma_{j}^{-2}\widetilde{Y}_{ij}\widehat{\boldsymbol{B}}_{j}^{\mathrm{T}}\mathrm{diag}(\boldsymbol{X}_{i})\widehat{\mathrm{E}}\boldsymbol{b}_{i}\end{array}\Big)^\mathrm{T},
\]
where $\widehat{\mathrm{E}}$ is the conditional expectation given the observed data, evaluated at the current estimates, $\widehat{\boldsymbol{B}}_j^{\mathrm{T}}$ is the $j$th row of $\widehat{\boldsymbol{B}}$, and $\boldsymbol{\Sigma}$ is evaluated at the current estimate. We update
\begin{align}
\left(\begin{array}{c}
\widehat{\boldsymbol{\mu}}\\
\widehat{\tau}
\end{array}\right)=\boldsymbol{Q}^{-1}\boldsymbol{u}.\label{eq:mutau}
\end{align}
Then, we update $\sigma^2_j$ ($j=1,\ldots,q$) using the closed-form expression:
\begin{align}
\widehat{\sigma}^2_j=\frac{1}{n}\sum_{i=1}^n\widehat{\mathrm{E}}\Big(\widetilde{Y}_{ik}-\mu_k-\tau\sum_{j=1}^pX_{ij}\widehat{\beta}_{jk}b_{ij}\Big)^2,\label{eq:sigma}
\end{align}
where $\boldsymbol{\mu}$ and $\tau$ are evaluated at the current estimates.

Note that the M-step involves only the first and second moments of the conditional distribution of $\boldsymbol{b}_i$. Conditional on $(\widetilde{\boldsymbol{Y}}_i,\boldsymbol{X}_i)$, $\boldsymbol{b}_i$ follows a normal distribution with mean
\begin{align}
\tau\mathrm{diag}(\boldsymbol{X}_i)\widehat{\boldsymbol{B}}^{\mathrm{T}}\big\{\tau^2\widehat{\boldsymbol{B}}\mathrm{diag}(\boldsymbol{X}_i)\widehat{\boldsymbol{B}}^{\mathrm{T}}+\boldsymbol{\Sigma}\big\}^{-1}(\widetilde{\boldsymbol{Y}}_i-\boldsymbol{\mu})\label{eq:mean}
\end{align}
and variance
\begin{align}
\boldsymbol{I}_p-\tau^2\mathrm{diag}(\boldsymbol{X}_i)\widehat{\boldsymbol{B}}^{\mathrm{T}}\big\{\tau^2\widehat{\boldsymbol{B}}\mathrm{diag}(\boldsymbol{X}_i)\widehat{\boldsymbol{B}}^{\mathrm{T}}+\boldsymbol{\Sigma}\big\}^{-1}\widehat{\boldsymbol{B}}\mathrm{diag}(\boldsymbol{X}_i),\label{eq:var}
\end{align}
where $\widetilde{\boldsymbol{Y}}_i=(\widetilde{Y}_{i1},\ldots,\widetilde{Y}_{iq})^{\mathrm{T}}$. Therefore, all expectations of functions of $\boldsymbol{b}_i$ in the E-step have closed-form expressions. We iterate between the E-step and the M-step until convergence.

The whole estimation algorithm can be summarized as follows:
\begin{enumerate}
\item[1.] Estimate $\boldsymbol{\xi}$ by maximizing $\widetilde{L}(\boldsymbol{\xi})$, with $\beta_{jk}$ fixed at $0$ for $m(j,k)=\infty$. Let $\widetilde{\beta}_{jk}$ be the estimated value of $\beta_{jk}$ in this initial estimation step.
\item[2.] Let $w_{jk}=|\widetilde{\beta}_{jk}|^{-1}$, and estimate $\boldsymbol{\xi}$ by maximizing $p\ell(\boldsymbol{\xi})$.
\item[3.] Refit the model selected in Step 2 by maximizing $\widetilde{L}(\boldsymbol{\xi})$, where parameters with zero estimates obtained from Step 2 are fixed at zero. Let $\widehat{\boldsymbol{B}}$ be the refitted estimate of $\boldsymbol{B}$.
\item[4.] Initialize $\tau$, and set $(\boldsymbol{\mu},\boldsymbol{\Sigma})$ to be the estimated values from Step 3. Compute $(\widehat{\tau},\widehat{\boldsymbol{\mu}},\widehat{\boldsymbol{\Sigma}})$, the maximizer of $L(\tau,\boldsymbol{\mu},\boldsymbol{\Sigma};\widehat{\boldsymbol{B}})$, by repeating the following E-step and M-step until convergence:
\begin{enumerate}
\item[a.] At the current estimates, compute $\widehat{\mathrm{E}}\boldsymbol{b}_i$ and $\widehat{\mathrm{E}}\boldsymbol{b}_i\boldsymbol{b}_i^{\mathrm{T}}$ from (\ref{eq:mean}) and (\ref{eq:var}).
\item[b.] Update $(\boldsymbol{\mu},\tau)$ using (\ref{eq:mutau}) with $\boldsymbol{\Sigma}$ set at the current estimate. Then, at the updated estimates of $(\boldsymbol{\mu},\tau)$, update $\boldsymbol{\Sigma}$ using (\ref{eq:sigma}).
\end{enumerate}    
\end{enumerate}

We propose to use a generalized information criterion to select the tuning parameters $\lambda$ and $\theta$. We suggest to select a small grid of values for $\theta$, and for each value of $\theta$ over the grid, we construct a grid for $\lambda$ using an approach similar to that of \cite{friedman2010regularization}. Let $\lambda_{\max,\theta}$ be the value of $\lambda$ such that the $\boldsymbol{B}$-component of $\arg\max_{\boldsymbol{\xi}}p\ell(\boldsymbol{\xi})$ (at the given $\theta$) is zero, and let $r$ be a small number, say 0.0001. Then, we set the grid of $\lambda$ to be $\lambda_{\max,\theta} r^{k/(K-1)}$ for $k=0,\ldots,K-1$, where $K$ is the size of the grid. For each value of $(\lambda,\theta)$ over the grid, compute the following generalized information criterion (GIC):
\[
\mathrm{GIC}(\lambda,\theta)=-2\log L(\widehat{\tau},\widehat{\boldsymbol{\mu}},\widehat{\boldsymbol{\Sigma}};\widehat{\boldsymbol{B}}) + \log(n)\log\{\log(p)\}\Vert\widehat{\boldsymbol{B}}\Vert_0,
\]
where the parameter estimates are computed under the tuning parameter values $(\lambda,\theta)$. The penalty term for model complexity follows that of \cite{wang2009shrinkage}. We select the set of tuning parameters $(\lambda,\theta)$ that minimizes $\mathrm{GIC}(\lambda,\theta)$.

From the estimated model, we can characterize the subject-specific activity of each mutation. For the $i$th subject, if the $j$th gene is mutated, i.e., $X_{ij}=1$, then we define the mutation score as
\[
\widehat{\mathrm{E}}\big\{(1+\widehat{\tau} b_{ij})^2\mid \widetilde{Y}_{i1},\ldots,\widetilde{Y}_{iq}\big\}\sum_{k=1}^q\widehat{\beta}_{jk}^2.
\]
This is the conditional expectation of the squared-$L_2$ norm of the effects of the $j$th mutation for the $i$th subject. Note that this quantity is a function of the first and second conditional moments of $\boldsymbol{b}_i$ and can be easily computed from the outputs of the EM algorithm. In addition, we can extend the definition of the mutation score to non-mutated genes. If $X_{ij}=0$, then we define the mutation score using the same expression as the above, but with $X_{ij}$ modified to be 1 in expressions (\ref{eq:mean}) and (\ref{eq:var}) in the computation of the conditional moments; the value of $X_{ij}$ in $\widetilde{\boldsymbol{Y}}_i$ remains unchanged.

\section{Simulation studies}\label{sec:simul}
We generated simulation data sets based on the METABRIC data that we analyze in Section 4. We set $p=91$ and $q=8939$, and the mutations and gene expressions correspond to the genes in the analysis of the METABRIC data. We also used the gene network adopted in the METABRIC data analysis. For a pair of genes $(j,k)$ with $m(j,k)>5$, we set $m(j,k)=\infty$. We independently generated each $X_j$ from the Bernoulli distribution, with the success probability set to be the maximum between 0.05 and the empirical mutation proportion of the gene in the METABRIC data set. We set 10 of the significantly mutated genes, namely PIK3CA, GATA3, MAP3K1, CDH1, TBX3, CBFB, RYR2, USH2A, AKT1, and NCOR1, to have effects on their downstream expressions. In particular, each mutation has effect $+1.5/-1.5$ on the expression of the same gene, where the sign is positive or negative with probability 0.5. Also, for each mutation, 5 genes with distance 1 from the mutation are randomly selected to have effect $+1/-1$, and 4 genes with distance 2 are randomly selected to have effect $+0.5/-0.5$; if the genes have fewer than 4/5 downstreams with the specified distance, then all the downstreams were selected to have effects. We set $\mu_k=0$ and $\sigma_k=1$ for $k=1,\ldots,q$ and set $\tau=0.5$. We set the sample size to be $n=1000$. We considered 100 simulation replicates.

Over the simulation replicates, the average number of nonzero elements of $\widehat{\boldsymbol{B}}$ is 240.2, and on average 74.56 of the nonzero elements correspond to a nonzero mutation effect; the total number of true nonzero elements of $\boldsymbol{B}$ is 80. The averaged value of $\widehat{\tau}$ is 0.438. Because we tend to select more nonzero elements of $\boldsymbol{B}$ than the true number of nonzero elements, a slight under-estimation of $\tau$ is expected. For the nonzero regression parameters, we present the empirical bias of the estimates in Table \ref{tab:sim1}.

\begin{table}
\protect\caption{Simulation results --- Empirical bias for nonzero parameters \label{tab:sim1}}
\centering
\renewcommand{\tabcolsep}{3pt}
\renewcommand{\arraystretch}{0.7}

\begin{threeparttable}
\begin{tabular}{rrrrrrrrrr}
\hline 
\multicolumn{2}{c}{AKT1} & \multicolumn{2}{c}{CBFB} & \multicolumn{2}{c}{CDH1} & \multicolumn{2}{c}{GATA3} & \multicolumn{2}{c}{MAP3K1}\tabularnewline
\hline 
True & Bias & True & Bias & True & Bias & True & Bias & True & Bias\tabularnewline
\hline 
$-$1.500 & 0.010 & 1.000 & $-$0.010 & 1.500 & $-$0.017 & 1.500 & 0.009 & $-$1.500 & 0.001\tabularnewline
$-$0.500 & 0.158 & $-$1.000 & 0.001 & $-$1.000 & $-$0.003 & $-$1.000 & 0.005 & 0.500 & $-$0.010\tabularnewline
$-$1.000 & $-$0.007 & 1.500 & 0.010 & $-$0.500 & 0.023 & 1.000 & $-$0.007 & $-$1.000 & $-$0.003\tabularnewline
$-$0.500 & 0.076 & $-$1.000 & $-$0.015 & 1.000 & $-$0.019 & $-$1.000 & $-$0.008 & $-$1.000 & 0.003\tabularnewline
$-$0.500 & 0.114 & 1.000 & $-$0.014 & 1.000 & $-$0.015 & $-$1.000 & $-$0.008 & 1.000 & 0.010\tabularnewline
$-$1.000 & $-$0.020 & 1.000 & $-$0.024 & $-$0.500 & 0.019 & 0.500 & $-$0.008 & $-$1.000 & 0.018\tabularnewline
1.000 & $-$0.006 & 0.500 & $-$0.079 & 0.500 & $-$0.024 & $-$0.500 & $-$0.010 & 1.000 & $-$0.008\tabularnewline
$-$1.000 & 0.007 & 0.500 & $-$0.063 & $-$1.000 & 0.002 & 1.000 & 0.015 & 0.500 & 0.001\tabularnewline
1.000 & 0.001 & 0.500 & $-$0.107 & 1.000 & $-$0.006 & $-$0.500 & $-$0.004 & $-$0.500 & 0.025\tabularnewline
0.500 & $-$0.177 & 0.500 & $-$0.061 & 0.500 & $-$0.033 & 0.500 & $-$0.012 & 0.500 & 0.000\tabularnewline
\hline 
\multicolumn{2}{c}{PIK3CA} & \multicolumn{2}{c}{RYR2} & \multicolumn{2}{c}{TBX3} & \multicolumn{2}{c}{USH2A} & \multicolumn{2}{c}{NCOR1}\tabularnewline
\hline 
True & Bias & True & Bias & True & Bias & True & Bias & True & Bias\tabularnewline
\hline 
0.500 & 0.011 & 1.500 & $-$0.008 & 1.500 & 0.009 & $-$1.500 & 0.004 & $-$0.500 & 0.114\tabularnewline
0.500 & 0.006 & 1.000 & $-$0.006 &  &  &  &  & $-$1.000 & $-$0.036\tabularnewline
1.500 & 0.007 & 1.000 & 0.011 &  &  &  &  & $-$0.500 & 0.122\tabularnewline
1.000 & 0.001 & $-$1.000 & 0.021 &  &  &  &  & 1.000 & 0.024\tabularnewline
0.500 & $-$0.003 & 0.500 & $-$0.099 &  &  &  &  & 0.500 & $-$0.174\tabularnewline
1.000 & 0.002 & $-$0.500 & 0.076 &  &  &  &  & $-$1.000 & 0.036\tabularnewline
0.500 & $-$0.009 & 0.500 & $-$0.069 &  &  &  &  & $-$0.500 & 0.202\tabularnewline
$-$1.000 & $-$0.003 & 0.500 & $-$0.133 &  &  &  &  & $-$1.500 & $-$0.004\tabularnewline
1.000 & 0.000 &  &  &  &  &  &  & 1.000 & $-$0.014\tabularnewline
$-$1.000 & $-$0.004 &  &  &  &  &  &  & 1.000 & 0.023\tabularnewline
\hline 
\end{tabular}
\begin{tablenotes}
 \linespread{1.2}
\item\small{}NOTE: Each row represents a nonzero effect of a mutation on a gene expression. Each pair of columns corresponds to the nonzero effects of a mutation on downstream expressions, ``True'' represents the true parameter value, and ``Bias'' represents the empirical bias. Because each mutation affects a different set of genes, for simplicity, we do not present the names of the downstream genes affected by the mutations.
\end{tablenotes}
\end{threeparttable}
\par
\protect
\end{table}

For each mutation with nonzero effects, we present the mean-squared error (MSE), defined as $\sum_{k=1}^q(\widehat{\beta}_{jk}-\beta_{0jk})^2$ for the $j$th mutation, the false discovery rate (FDR), the true positive rate (TPR), and the sample correlation (Corr) between the estimated mutation score and the true mutation score, where $\beta_{0jk}$ denotes the true value of $\beta_{jk}$. For the correlation between the estimated and true mutation scores, the true mutation score for the $j$th mutation is defined as $(1+\tau b_{ij})^2\sum_{k=1}^q\beta_{0jk}^2$, and the sample correlation is computed using only the subjects with the mutation. The results are presented in Table \ref{tab:sim2}. In the table, we also present the number of downstreams under the gene network for these 10 mutations.

\begin{table}
\protect\caption{Simulation results --- point estimation, variable selection, and mutation score prediction
\label{tab:sim2}}
\centering
\renewcommand{\tabcolsep}{5pt}
\renewcommand{\arraystretch}{0.7}

\begin{threeparttable}
\begin{tabular}{cccccc}
\hline 
 & \# Downstreams & MSE & FDR & TPR & Corr\tabularnewline
\hline 
AKT1 & 6828 & 2.061 & 0.432 & 0.860 & 0.771\tabularnewline
CBFB & 6689 & 0.743 & 0.137 & 0.906 & 0.787\tabularnewline
CDH1 & 6761 & 0.682 & 0.287 & 0.977 & 0.787\tabularnewline
GATA3 & 6700 & 0.256 & 0.152 & 0.999 & 0.799\tabularnewline
MAP3K1 & 6758 & 0.444 & 0.229 & 0.991 & 0.792\tabularnewline
PIK3CA & 6795 & 0.272 & 0.306 & 1.000 & 0.797\tabularnewline
RYR2 & 1828 & 0.406 & 0.003 & 0.864 & 0.752\tabularnewline
TBX3 & 1 & 0.028 & 0.000 & 1.000 & 0.583\tabularnewline
USH2A & 1 & 0.029 & 0.000 & 1.000 & 0.569\tabularnewline
NCOR1 & 6723 & 1.292 & 0.211 & 0.832 & 0.774\tabularnewline
\hline 
\end{tabular}
\begin{tablenotes}
 \linespread{1.2}
\item\small{}
\end{tablenotes}
\end{threeparttable}
\par
\protect
\end{table}

The biases of the parameter estimators are overall small, suggesting that the pseudo-likelihood approach yields consistent estimation. For some parameters, especially those with true value of 0.5 or $-$0.5, the bias may be large (with a magnitude up to 0.2). This is because the corresponding parameters are not selected in some replicates, thus yielding a zero point estimate, so the averaged estimates are biased towards zero. 

For all mutations with nonzero effects, the MSE are small, especially in light of the large number of downstream genes. The FDR are mostly around 0.2--0.3, and the three genes with particularly few downstream genes have very small FDR. The TPR are all larger than 0.8. With the exception of TBX3 and USH2A, the correlations for mutation score are moderately large, with values between 0.75 and 0.8. The genes TBX3 and USH2A have only one downstream gene each. Therefore, although the point estimates for the mutation effects are accurate, the information about the random effect $b_{ij}$ contained in the observed data is relatively small, so the correlations for mutation scores are also small.

\section{Real data analysis}
We analyzed a data set of breast cancer patients of Luminal A and Luminal B subtypes from the METABRIC study \citep{curtis2012genomic,pereira2016somatic}. The sample size is 959, with 515 of subtype Luminal A and 444 of subtype Luminal B. The whole data set contains 25186 gene expressions, and $q=8939$ genes have available network information. For mutation data, we only keep genes with mutation frequency larger than or equal to 10, resulting in the number of mutations of $p=91$.

We adopted the gene network used in \cite{siegel2018integrated}, which was built from both curated and non-curated human gene interactions obtained from \cite{ciriello2012mutual} and \cite{kanehisa2012kegg}. In the network, the numbers of pairs of $(j,k)$ with $m(j,k)=$ 0, 1, 2, 3, 4, and 5 are 91, 3932, 49690, 184655, 179181, 71325, respectively.

We estimate the model using the proposed two-step approach. We considered tuning parameter values of $\theta=(1+e)^{-1}$, 0.5, and $e(1+e)^{-1}$ and selected the tuning parameters using the proposed GIC. As in the simulation studies, we set $m(j,k)=\infty$ for $m(j,k)>5$. In the estimated model, the total number of nonzero $\widehat{\beta}_{jk}$'s is 10227. The number of nonzero $\widehat{\beta}_{jk}$'s with $m(j,k)=$ 0, 1, 2, 3, 4, and 5 are 26, 428, 3185, 4904, 1458, and 226, respectively. Therefore, the sparsity of $\widehat{\beta}_{jk}$ grows as $m(j,k)$ increases. Of all the mutations, 80 of them are estimated to have nonzero effects on at least one gene expression. PIK3CA, TP53, and CDH1 have effects on the largest numbers of genes expressions, and they affect 2227, 1716, and 1018 gene expressions, respectively. Among the 91 mutations, the median number of expressions affected is 28. The estimated value of $\tau$ is 0.688, suggesting a moderate degree of heterogeneity in mutation effects.

\begin{figure}
\begin{center}
\includegraphics[scale=0.45]{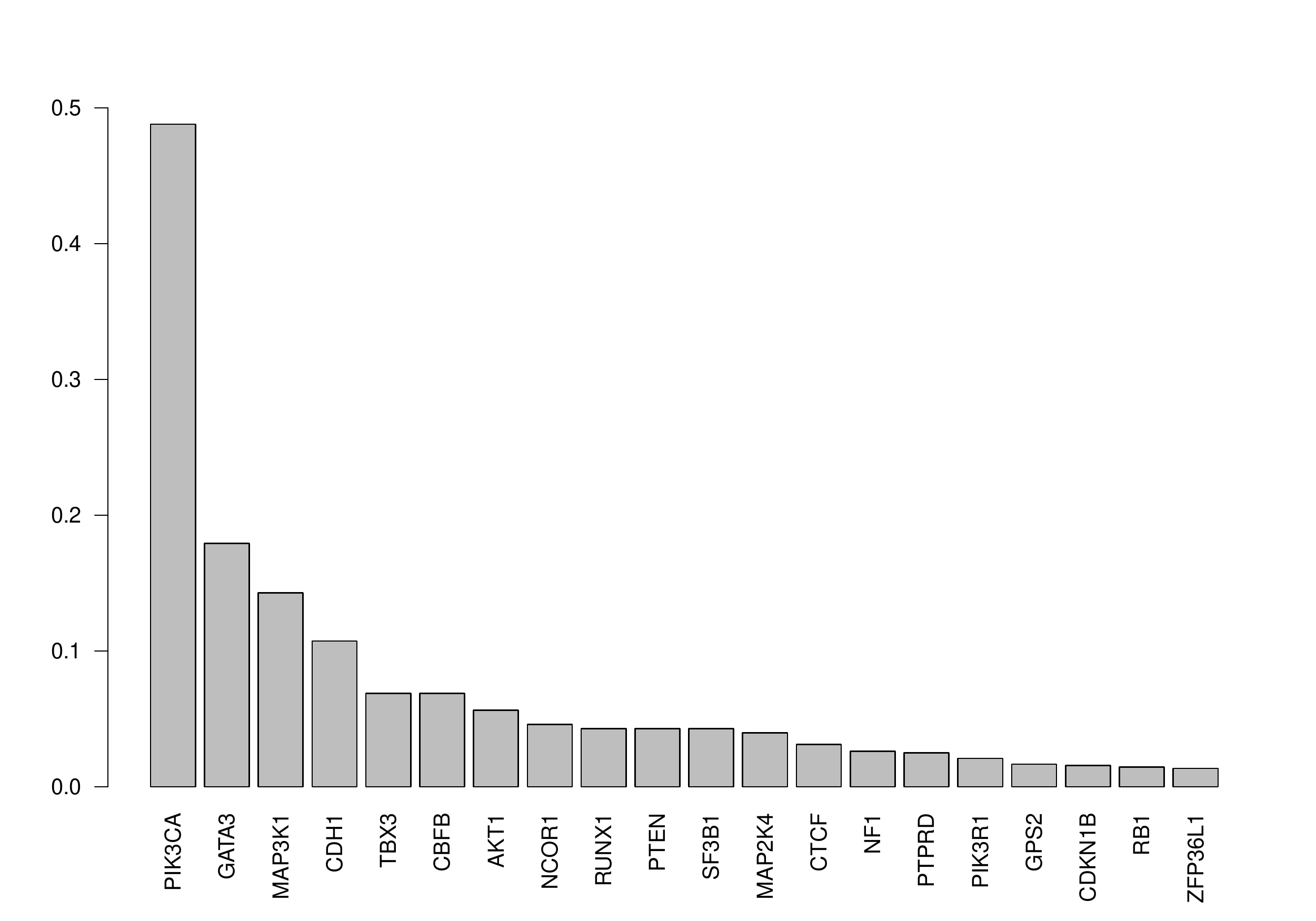}
\caption{Mutation rates for the SMG}\label{fig:mutrate}
\end{center}
\end{figure}

To evaluate the biological significance of the estimated model and especially the mutation scores, we focus on a set of mutations that have been reported to be relevant to breast cancer. In particular, we consider the significantly mutated genes (SMG) list in \cite{cancer2012comprehensive} and take a subset of 20 genes that are estimated to have effects on some gene expressions. Figure \ref{fig:mutrate} shows the mutation rates of these genes. The mutation rate is about 0.5 for PIK3CA, whereas those for the other genes are lower than 0.2.

\begin{figure}
\begin{center}
\includegraphics[scale=0.5]{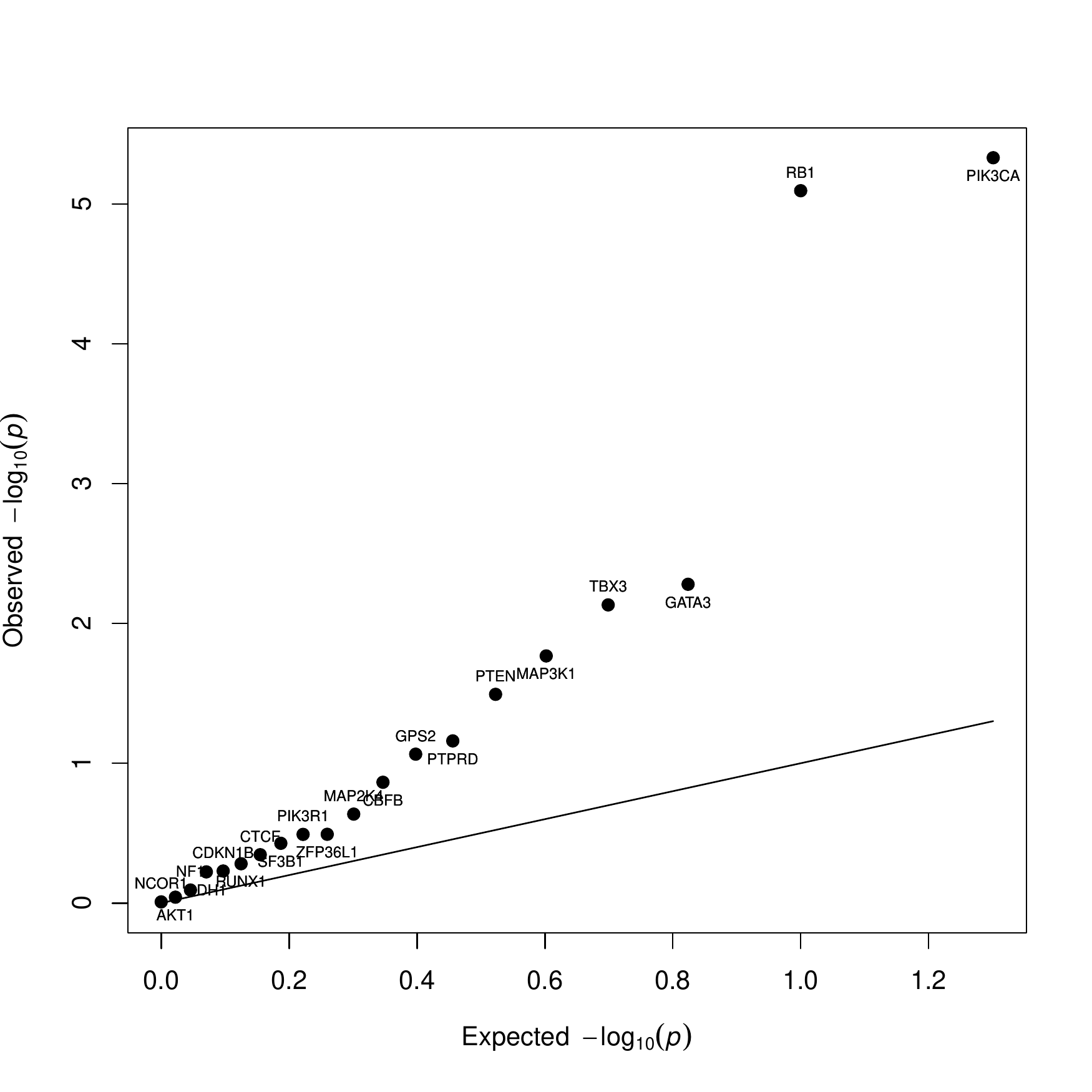}
\caption{QQ-plot for the logrank test $p$-values for survival VS mutation score}\label{fig:logrank}
\end{center}
\end{figure}

\begin{figure}
\begin{center}
\includegraphics[scale=0.5]{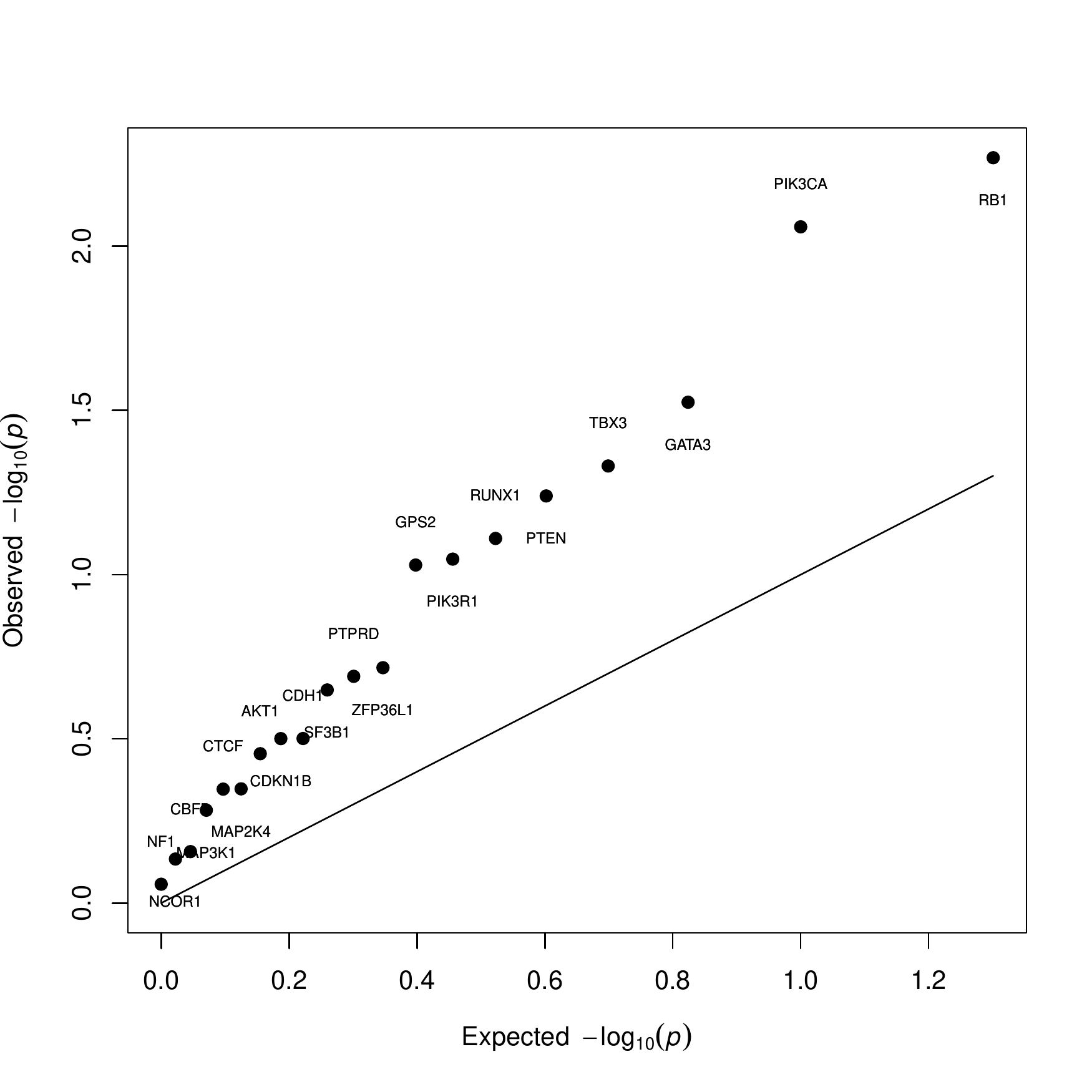}
\caption{QQ-plot for the Cox $p$-values for survival VS mutation score, accounting for subtype}\label{fig:cox}
\end{center}
\end{figure}

We calculate the mutation scores for these 20 mutations for all subjects. To assess the biological relevance of the mutation score, we study the association between the score and survival time. We consider the time to death due to disease, treating death by other causes or lost-to-followup as right censoring. The median follow-up time is 124 months, and the censoring proportion is 72.9\%. For each of the 20 SMG, we classify subjects with that mutation into high-score group versus low-score group, according to whether the score is larger than or smaller than the median score for the gene among the mutated subjects; we similarly divide subjects into high-score and low-score groups for subjects without the mutation. Then, we perform the logrank test for the survival time between the high-score and low-score groups (regardless of the mutation status). To account for cancer subtypes, we also fit the Cox model of the survival time against the score status and subtype indicator. Figure \ref{fig:logrank} shows the QQ-plot for the $p$-values of the logrank tests, and Figure \ref{fig:cox} shows the QQ-plot for the Wald test $p$-values for the score status under the Cox model.

With or without accounting for subtype, the associations between mutation score and survival time are stronger than that expected under the null hypothesis. The results suggest that for these known important genes, the scores, as summaries of gene expressions, are associated with the survival time, and the associations remain even after accounting for subtype.

We further investigate the association between the gene score and survival time for the two most frequently mutated genes in the data set, PIK3CA and GATA3.

\subsubsection*{PIK3CA}
\begin{figure}
\begin{center}
\includegraphics[scale=0.5]{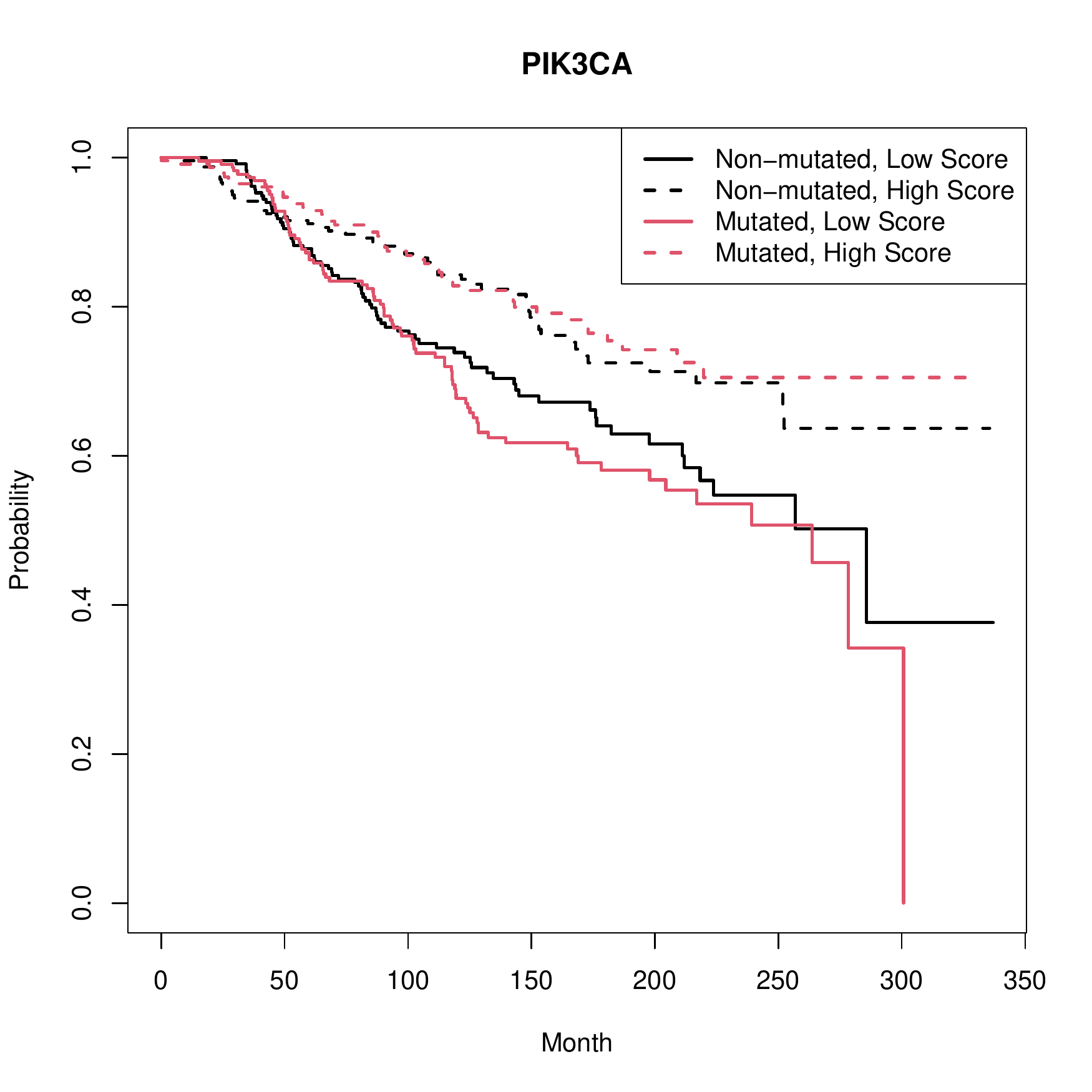}
\caption{KM plots for subjects with different PIK3CA mutation statuses and score levels}\label{fig:PIK3CAscore}
\end{center}
\end{figure}

Figure \ref{fig:PIK3CAscore} shows the Kaplan--Meier curves for the high-score and low-score groups of subjects with or without a mutation in PIK3CA.
In both the mutation and non-mutation groups, subjects with high scores tend to survive longer than subjects with low scores. The improvement in survival is more substantial in the mutation group (logrank $p$-value of $7.12\times10^{-5}$) than in the non-mutation group (logrank $p$-value of 0.0110). This suggests that the gene expressions identified to be associated with the mutation of PIK3CA, as summarized by the gene score, are relevant to the survival time. Remarkably, if we compare the survival times between the mutated and non-mutated subjects, there is no significant difference between the survival functions of the two groups (logrank $p$-value $=0.739$). This suggests that the mutation status alone, without consideration of the expression of genes associated with the mutation, does not effectively distinguish subjects with different survival patterns.

\begin{figure}
\begin{center}
\includegraphics[scale=0.5]{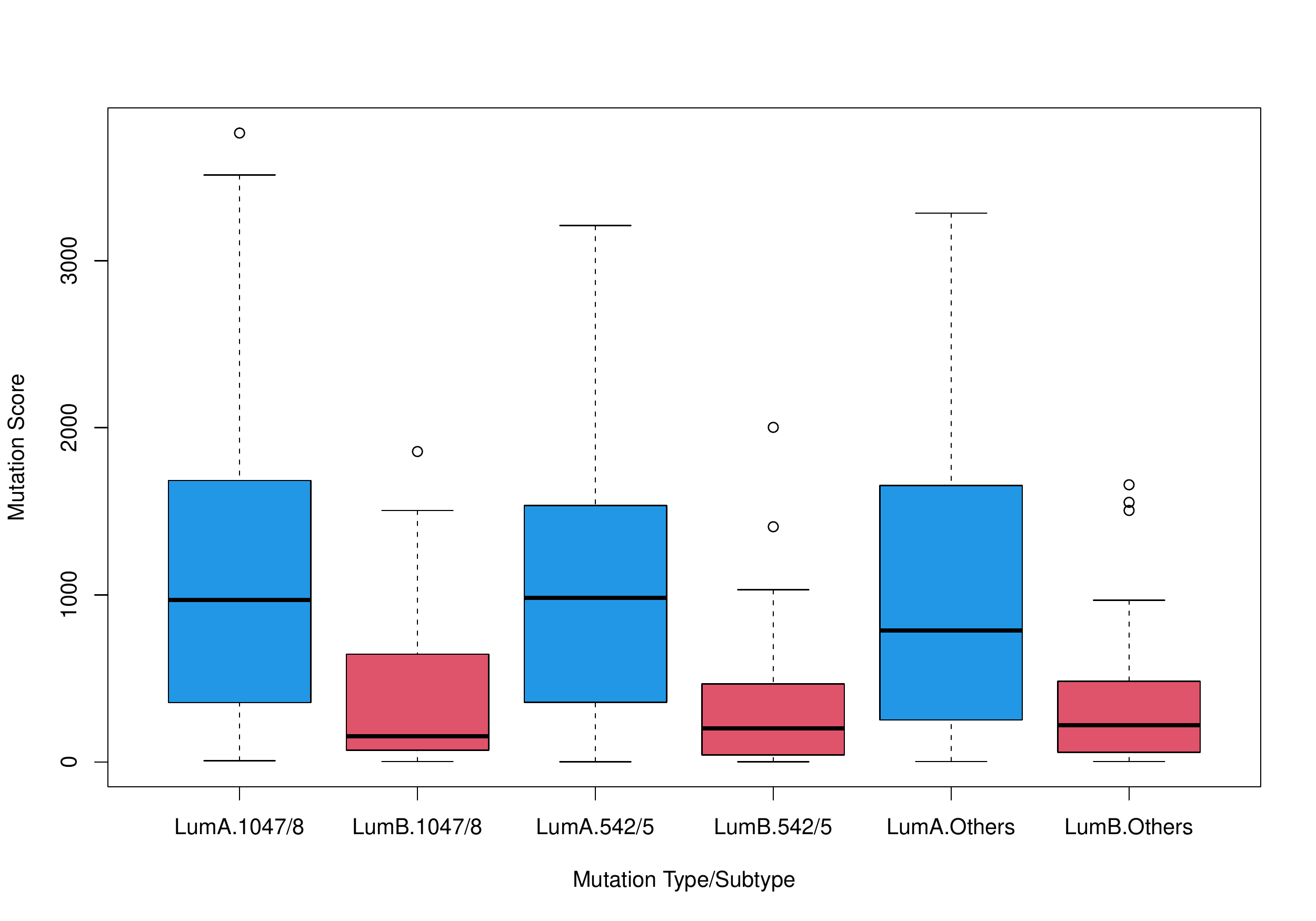}
\caption{PIK3CA score level for subjects with different subtypes and mutation types}\label{fig:PIK3CAscoreVStype}
\end{center}
\end{figure}

We evaluate the association between mutation score and the mutation type. In the regression analyses, we treat a gene as mutated if it harbours any non-silent mutation within its genetic region. In the data set, the location of the mutation is actually available. Here, we classify the mutation into three groups: those located at 1047/1048, those located at 542/545, and others. Figure \ref{fig:PIK3CAscoreVStype} shows the boxplot of mutation score over types of mutations and subtypes. There is no substantial difference in score between the three groups of mutation, but the difference is substantial between the two subtypes. In this case, the gene score captures the difference between the two subtypes.

\subsubsection*{GATA3}
\begin{figure}
\begin{center}
\includegraphics[scale=0.5]{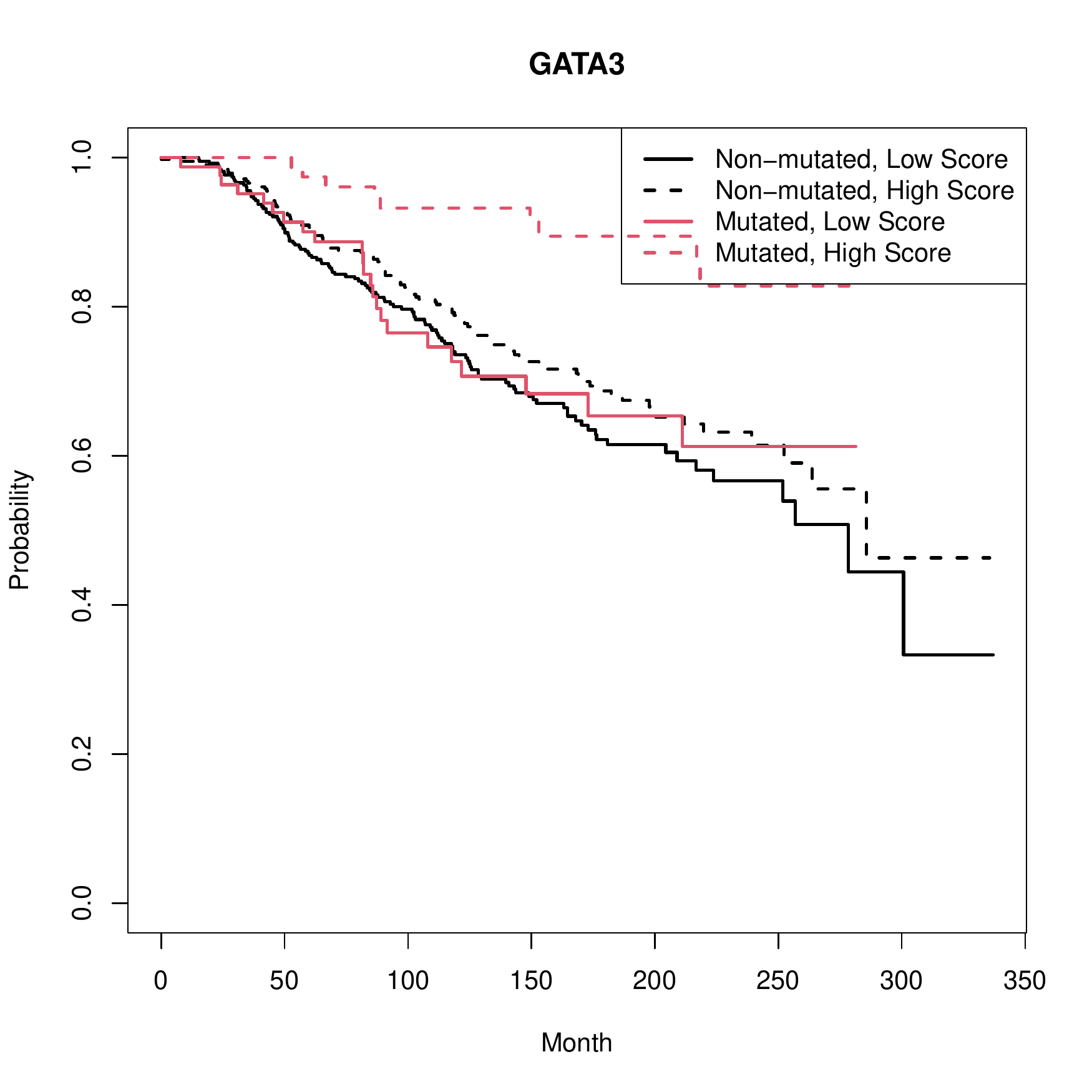}
\caption{KM plots for subjects with different GATA3 mutation statuses and score levels}\label{fig:GATA3score}
\end{center}
\end{figure}

\begin{figure}
\begin{center}
\includegraphics[scale=0.5]{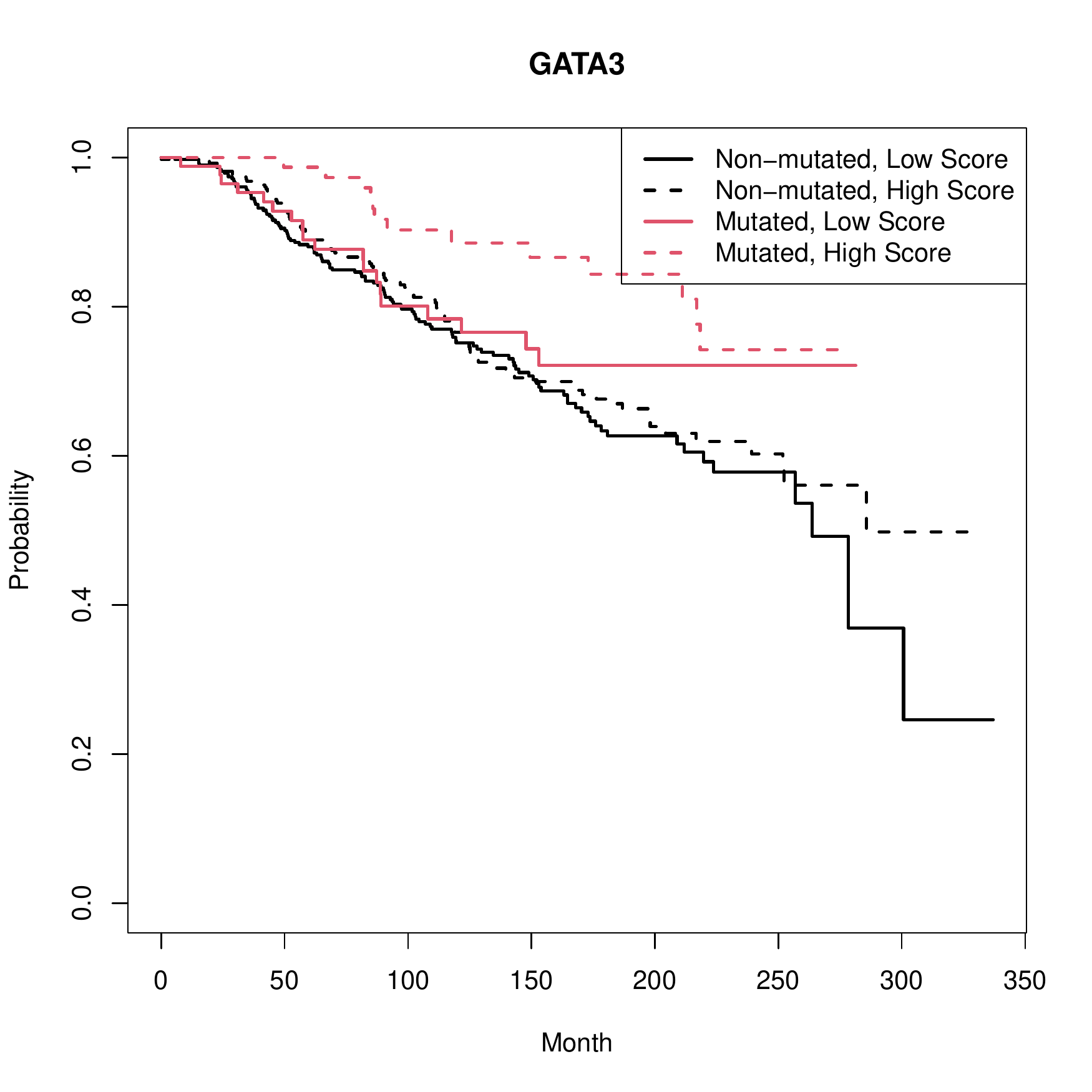}
\caption{KM plots for subjects with different GATA3 mutation statuses and gene expression levels}\label{fig:KM-GATA3-expression}
\end{center}
\end{figure}

For GATA3, the Kaplan--Meier curves for both mutation groups and high/low gene score groups are plotted in Figure \ref{fig:GATA3score}.
For both the mutation and non-mutation groups, subjects with high scores have overall longer survival time. The logrank test for the survival difference between the high-score and low-score groups yields a $p$-value of 0.00525. Remarkably, the two groups with no mutation and the group with mutation but low mutation score have similar survival distributions, whereas the group with mutation and high mutation score has substantially longer survival time. This suggests that it is the combination of mutation and perturbation of the associated gene expressions that drive the biological difference among the cancer patients, whereas the mutation alone may not be as effective. We perform a similar comparison of survival time but with the expression of GATA3 instead of the mutation score, and the curves are plotted in Figure \ref{fig:KM-GATA3-expression}. The group with mutation and high expression of GATA3 has longer survival than the remaining groups, but the difference between groups with high and low expression values is no longer significant under the logrank test ($p$-value $=0.178$). This suggests that the mutation score captures the biological status of the subjects better than the expression of a single gene.

\begin{figure}
\begin{center}
\includegraphics[scale=0.5]{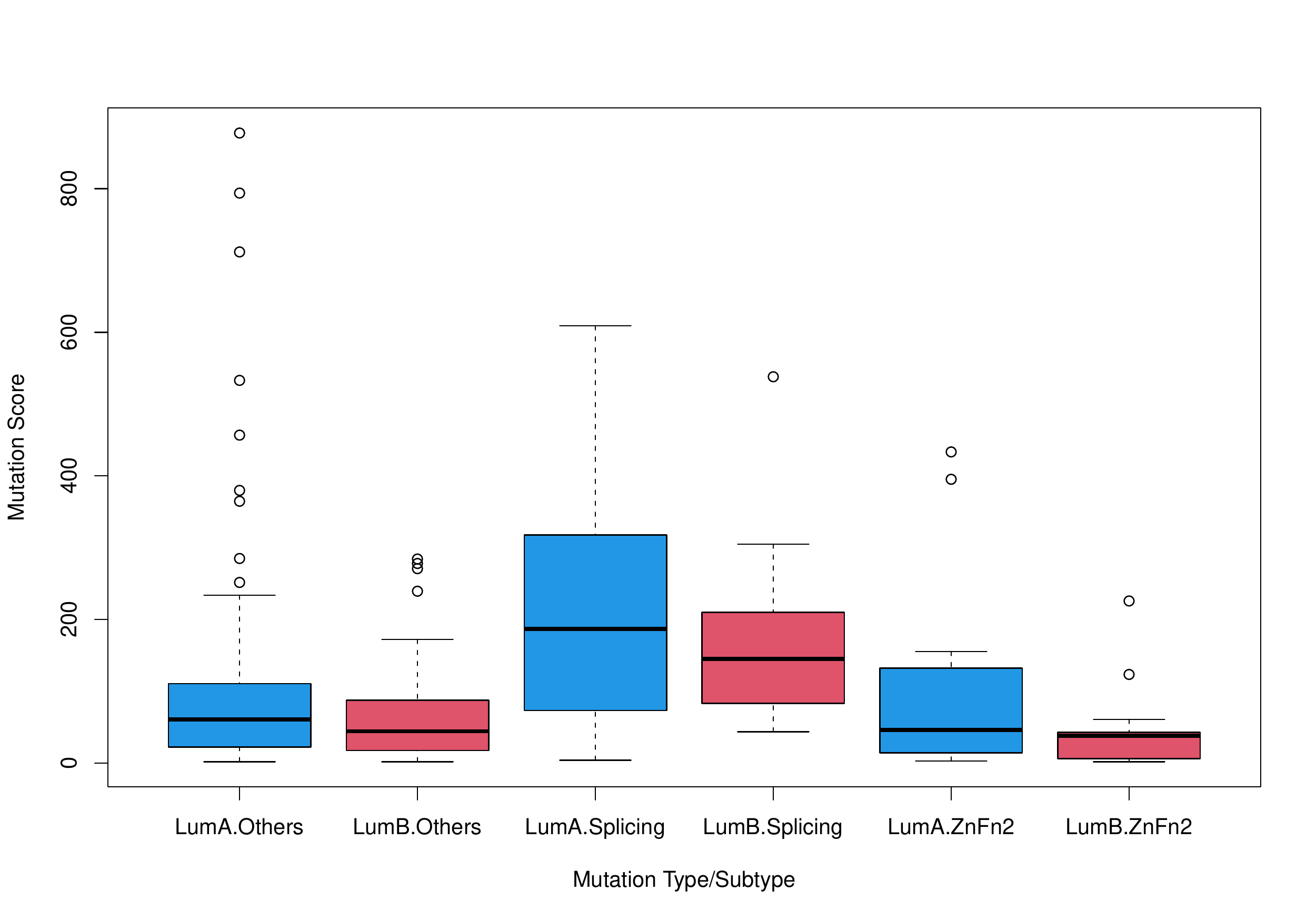}
\caption{GATA3 score level for subjects with different subtypes and mutation types}\label{fig:GATA3scoreVStype}
\end{center}
\end{figure}

\begin{figure}
\begin{center}
\includegraphics[scale=0.5]{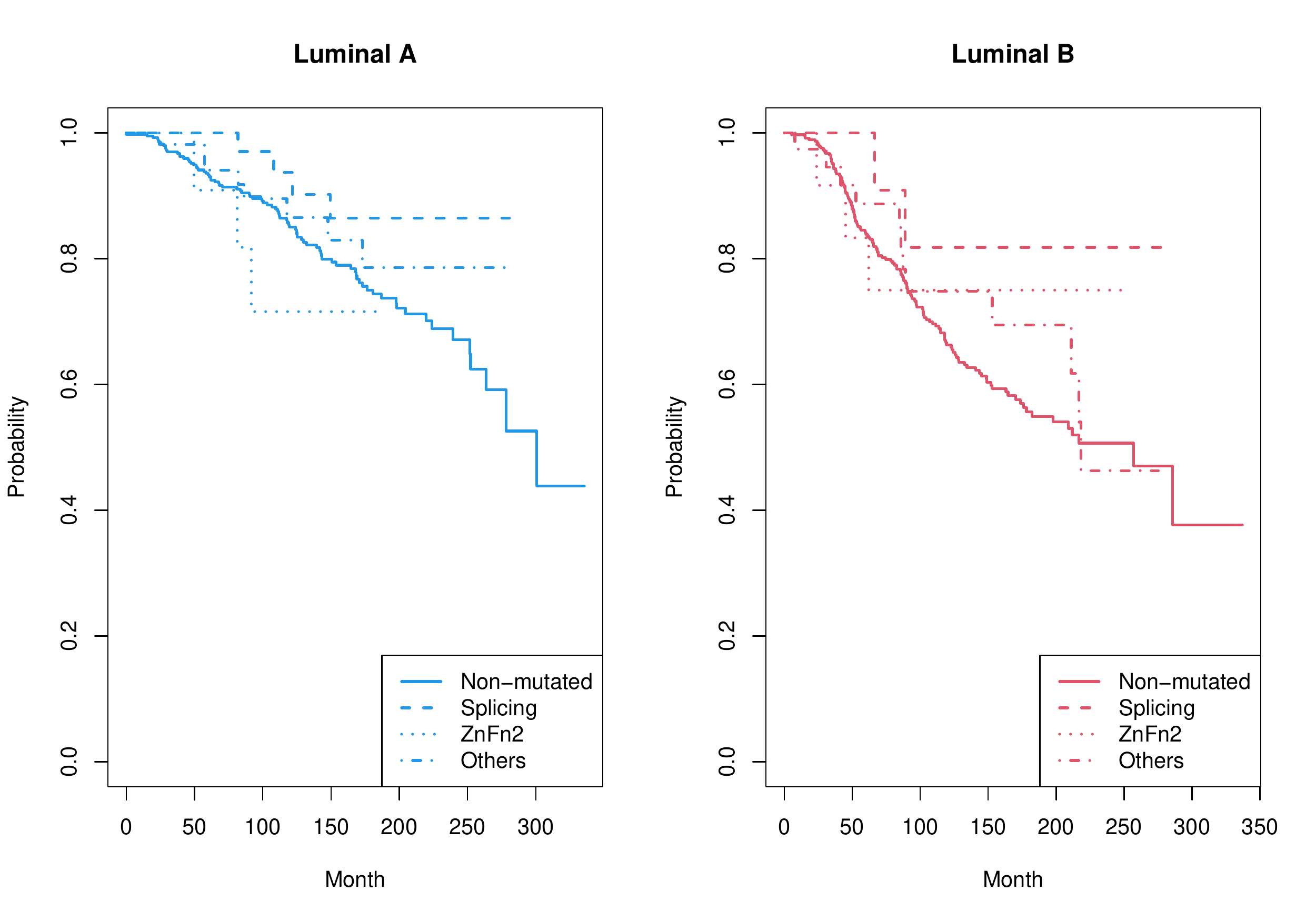}
\caption{KM plots for subjects with different subtypes and GATA3 mutation types}\label{fig:GATA3survOverType}
\end{center}
\end{figure}

We evaluate the distributions of the mutation score across different types of mutation. Following \cite{takaku2018gata3}, we classify the mutations into three groups: the second GATA3 zinc-finger (ZnFn2) mutation, splice site mutations, and others. We present the boxplot for the mutation score across the three mutation groups and subtypes in Figure \ref{fig:GATA3scoreVStype}. We can see that for both subtypes, the scores for splice site mutations tend to be higher than the other two groups. This could explain why the combination of high score and mutation leads to change in survival time, but mutation alone may not: it is the splice site mutations that lead to difference in the survival distribution. Figure \ref{fig:GATA3survOverType} shows the KM plots for subjects with different types of mutations. We can see that for both subtypes, subjects with splice site mutation tend to have the highest survival probabilities.

\section{Discussion}
There are two main steps in the computation of the mutation score --- model estimation and evaluation of the conditional expectation of the mutation effects. In principle, these two steps can be separated and performed on different data sets. A model can be estimated from a previous study or otherwise prespecified from prior knowledge, and then the mutation score can be computed for new subjects. Similar to DawnRank, the proposed score can be computed even for only a single subject with available mutation and gene expression data, as long as information
about the association between mutation and gene expressions of the population where this subject belongs to is known.

If a gene network is not available, then the estimation of $\boldsymbol{B}$ can be replaced by a standard multivariate lasso regression,
with a $(q\times p)$ unknown regression parameter matrix and a universal tuning parameter for all regression parameters.
The subsequent EM algorithm and score calculation can be performed exactly as proposed.
Alternatively, other penalization methods for the estimation of $\boldsymbol{B}$
that encourages group structures, such as \cite{zhu2016integrating}, can be adopted.

The proposed method assumes that for a mutation to have effect on the expression of a gene, there must exist a directed path from the mutation to
the gene in the network. To relax this assumption, we may set all mutation-to-expression effects as free parameters to be estimated and introduce another penalty for the effects that do not correspond to a path in the network.

\bibliographystyle{jasa}

\end{document}